# Upswing in Industrial Activity and Infant Mortality during Late 19th Century US

**Nahid Tavassoli**[1]✉
**Hamid Noghanibehambari**[2]
**Farzaneh Noghani**[3]
**Mostafa Toranji**[4]

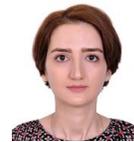

(✉ *Corresponding Author*)

[1] *PhD Student, Department of Economics, Texas Tech University, USA.*
*Email: nahid.tavassoli@ttu.edu Tel: +1-806-516-3929*
[2] *PhD Candidate, Department of Economics, Texas Tech University, USA.*
*Email: Hamid.Noghanibehambari@ttu.edu Tel: +1-806-620-1812*
[3] *PhD Candidate, Rawls Business School, Texas Tech University, USA.*
*Email: farzaneh.noghani@ttu.edu Tel: +1-806-252-3094*
[4] *Master Student, Department of Economics, University of Tehran, Iran.*
*Email: Mostafa.Toranji@ut.ac.ir Tel: +98-933-458-9134*

## Abstract

This paper aims to assess the effects of industrial pollution on infant mortality between the years 1850-1940 using full count decennial censuses. In this period, US economy experienced a tremendous rise in industrial activity with significant variation among different counties in absorbing manufacturing industries. Since manufacturing industries are shown to be the main source of pollution, we use the share of employment at the county level in this industry to proxy for space-time variation in industrial pollution. Since male embryos are more vulnerable to external stressors like pollution during prenatal development, they will face higher likelihood of fetal death. Therefore, we proxy infant mortality with different measures of gender ratio. We show that the upswing in industrial pollution during late nineteenth century and early twentieth century has led to an increase in infant mortality. The results are consistent and robust across different scenarios, measures for our proxies, and aggregation levels. We find that infants and more specifically male infants had paid the price of pollution during upswing in industrial growth at the dawn of the 20th century. Contemporary datasets are used to verify the validity of the proxies. Some policy implications are discussed.

**Citation** | Nahid Tavassoli; Hamid Noghanibehambari; Farzaneh Noghani; Mostafa Toranji (2020). Upswing in Industrial Activity and Infant Mortality during Late 19th Century US. Journal of Environments, 6(1): 1-13.
**History:**
Received: 28 May 2020
Revised: 30 June 2020
Accepted: 4 August 2020
Published: 21 August 2020
**Licensed:** This work is licensed under a Creative Commons Attribution 3.0 License
**Publisher:** Asian Online Journal Publishing Group

**Acknowledgement:** All authors contributed to the conception and design of the study.
**Funding:** This study received no specific financial support.
**Competing Interests:** The authors declare that they have no conflict of interests.
**Transparency:** The authors confirm that the manuscript is an honest, accurate, and transparent account of the study was reported; that no vital features of the study have been omitted; and that any discrepancies from the study as planned have been explained.
**Ethical:** This study follows all ethical practices during writing.

## Contents

**Contribution of this paper to the literature**
This study contributes to the existing literature by assessing the effects of industrial pollution on infant mortality between the years 1850-1940 using full count decennial censuses.

## 1. Introduction

It is widely documented that pollution has a strong effect on prenatal development, fetal death, infant mortality, infants health, and long-term human capital outcomes (refer to Currie (2011) for a review). Scarcity of data for both infant mortality and pollution measures has forced most of the quantitative research to be based on cross-sectional analyzes, smaller time periods in panel data, and using more recent years. One common source of exogenous variation in air pollution, prevalent in this research, is Clean Air Act Amendments of 1970 (CAAA) which is used to assess the effects of a reduction in air pollution on fetal death rates (Sanders & Stoecker, 2011) infants mortality (Chay & Greenstone, 2003a); (Chay & Greenstone, 2003b) the labor force participation and earnings at age 30 (Isen, Rossin-Slater, & Walker, 2017) high school test scores (Sanders, 2012) and absenteeism among schoolchildren (Ransom & Pope III, 1992).

In this paper, we try to find the association between air pollution and infant mortality during late $19^{th}$ century and the dawn of $20^{th}$ century. Studying this association in a historical period has two advantages. First, In this time period, it was much less likely for families to migrate from one county to another for carrier perspectives and the least for pollution reasons. Another advantage of focusing on this time span is the large upswing in industrial activities and pollutants. A mitigation in air pollutants as a result of CAAA or reductions caused by recessions (Chay & Greenstone, 2003b) might have asymmetric effects com- pared to the times when pregnant mothers and infants are susceptible to large increases in pollutants. Distinct variation among counties and over a considerably large period of time (1850–1940) in degrees of industrialization provides a rich panel data to assess the effect of pollution on fetal and infant deaths, with less concerns that migration issues bias the results. Going one and a half century back in time comes with a price. Neither fetal death nor any measure of pollution are available for the first half of $20^{th}$ century, let aside $19^{th}$ century. To overcome these challenges, we introduce some proxies for both dependent and independent variables.

Manufacturing industries has been among the most pollution intensive industries and for this reason it has been used as a proxy for pollution (Beach & Hanlon, 2018); (Sanders, 2012). Hence, we use employment in manufacturing, and the share of employment in manufacturing to total population as a proxy for pollution. Later, we add employment in construction and mining industries to manufacturing in order to construct some modified versions of these instruments.

Like pollution, data on infant mortality and more importantly fetal death has been scarce. Vital Statistics Data Files start reporting publicly available data on deaths and births from 1968. Prenatal and fetal deaths files are available for even later times (starting at 1995). Although some mortality reports is available for the first half of the 20th century but they are aggregated at state level, sporadic in time, and more likely suffer from selection issues. However, gender ratio has been shown to provide a quite robust proxy for fetal deaths. Male embryos are more susceptible to external stressors like pollution during prenatal development and more likely to suffer health problems during infancy as a respond to such stressors (Sanders. & Stoecker, 2011). Therefore, during periods of large external shocks to utero, the reproductive system will be biased to produce more XX–chromosomes combination rather than XY–combination and as a consequence the probability of a live birth being girl increases. The latter probability is the dependent variable in this paper for analysis at household level while the female ratio to total number of children is the proxy used for county level regressions.

Using these proxies, we show that one thousandth of logarithm of employment in manufacturing is associated with 1–3% increase in the probability of a live birth being female. We find that Mothers in counties with higher degrees of manufacturing industries were more likely to give birth to daughters, experience the fetal, prenatal, and infant death.

Although all the coefficients are relatively small, one fact highlights their importance. A 1% increase in the probability of having more daughters than sons, or in other words, a 1% increase in the ratio of female newborns to total livebirths is a strong respond from the reproductive systems of mothers to environmental shocks. As Fisher's Law (Fisher, 1999) states in a balanced environment the sex ratio of most species that produce offspring through.

sexual reproduction is 1:1. Since humans are labile species and their sex are determined at the time of the conception, the environmental effects must have played insignificant roles in their sex determination. It is, thus, the severity of external shocks from environment that causes the maternal reproductive system to generate skewed sex-determination distribution. A 1% deviation from the balanced path of sex determination is a red signal that the probability generator of mother is sending although inadvertently, not for the sake of a feedback but as a mechanism for increasing chances of species survival.

## 2. A Brief Literature Review

Literature on health, pollution and individual outcomes is extremely large. Isen et al. (2017) uses US administrative data to evaluate the long-term effects of 1970 Clean Air Amendment Act. It exploits the variation of differential exposures of counties which are affected by the act, and finds that higher exposures in the year of birth is associated with lower earnings and labor force participation in the age of 30. Chay and Greenstone (2003b) in their seminal work, uses variation in pollution exposures in US counties caused by 1981–1982 recession to estimate the effects of pollution on infant mortality rates and finds that 1% reduction in total suspended particulates (TSP) will cause a 0.35% decline in infant mortality. Some natural experiments have been applied as an exogenous shock to air pollution in order to evaluate the contemporaneous outcome of infant health.

Sanders (2012) attempts to link prenatal exposure to Total Suspended Particulates at county level to educational outcomes of children in their tenth grade measured as math and reading exams in Texas. They followed Chay and Greenstone (2003b) and used variation in timing and different degrees of reduction in TSPs among counties caused by 1980's recessions as the shock to exposure of mothers during pregnancy. One standard deviation decrease in TSPs in a student's year of birth had led to 2% increase in standard deviation in high school test scores. It then used relative manufacturing employment as an instrumental variable for pollution at county level and found that a

**2**

standard deviation decrease in share of manufacturing employment increased high school test scores by 6% of one standard deviation.

Beach and Hanlon (2018) estimates the health consequences of air pollution in 1851–60 in England. Due to lack of data on air quality and pollution, it uses local coal use level based on local industrial structure and industry–specific coal use intensity as an instrument for industrial pollution. It finds that one standard deviation increase in coal use raised infant mortality by 6–8%. To the best of my knowledge, an investigation of this matter has not been done with US data while it seems necessary. The most important reason is the lack of data. Not only there is no data on air pollution for US at the dawn of $20^{th}$ century but also there is no publicly available datasets on infant mortality or infant health before 1968, the beginning of Vital Statistics Data Files.

Inspiring from Sanders (2012) and Beach and Hanlon (2018) we proxy air pollution by share of employed people in manufacturing industries to the total population and absolute value of the size of manufacturing industries measured by employment at county level. More- over, we proxy infant mortality with gender ratio of live births as introduced in Sanders. and Stoecker (2011). Although the former proxy has been widely used in the literature, the latter is a completely new proxy and has a solid background in evolutionary biology (Trivers & Willard, 1973); (Wells, 2000);(Charnov, 1993); (Booksmythe, Gerber, Ebert, & Kokko, 2018). Using these two proxy variables, we intend to capture the effect of an increase in air pollution caused by growth in manufacturing industries on fetal death and infant mortality in US between 1850–1940.

## 3. Data Overview and Sample Selection

In this paper we use a variety of publicly available data sources. This section overviews our data sources as well as variable constructions and sample selections.

The main source of data is full count decennial censuses for the years 1850, 1880, 1900, 1910, 1920, 1930, and 1940. The integrated dataset has been extracted from (Ruggles, Genadek, Goeken, Grover, & Sobek, 2017). In the preferred scenario, the gender ratio of children under age 5 at household level or at county level are being computed to use as a metric for fetal death and infant mortality. The 1860 and 1870 data were not used since the full count was not publicly available and so for many counties the gender ratio of these children could not have been calculated. Moreover, the Civil War between 1861–1865 could have had biased estimates for the year 1870 in a way not only related to employment in manufacturing as the variable of interest. The 1890 census has not been available because almost all the relevant documents has been devastated during an overnight conflagration (Blake, 1996). All missing data on children, mother characteristics, and household head socioeconomic index are excluded.

We exclude all mothers for whom the age of eldest child is more than the eldest child residing in the household since for such cases the gender of all children is unavailable. As a consequence of such exclusions, all children who are born from the mother must be present at the household. Moreover, all adopted, stepchildren, and other relatives recorded as a child are excluded. Only children who are given births from the present mother at the household are being considered since adopted or stepchildren could have been born in other counties and brought in their current residence after birth. We try to link the household head to the mothers of children since during the whole time-span about 1 million households existed in which more than one mother had at least one own birth child. Since the dependent variable is the probability that a mother, whose eldest child's age does not surpass a certain limit, has more daughters than sons (if any) conditional on having at least one own birth child, all families for whom there is no child present at the household or the eldest child of the mother is above the threshold are excluded.

In Figure 2 counties are divided into terciles based on the number of residents in the county who are employed in manufacturing industries. The left panel shows the unprecedented increase in manufacturing industries for almost all years in the sample except 1860 and 1870, which can be attributed to the Civil War. In the right panel, trends of gender ratio (female to male) are illustrated. Regardless of the time trend, counties with high manufacturing employment had shown considerably higher gender ratios for the exception of 1910.

Figure 1 maps employment in manufacturing and female ratio of children under age 5 at county level for United States excluding its territories. Eastern counties have experienced more people working in manufacturing industries and meanwhile higher gender ratios for their newborns. Far Western counties depict the same narrated while most of southern and central counties are vice versa. It must be noted that the GIS files are withdrawn from NHGIS and linked to statistics computed using 1900 census file (data source: Manson, Schroeder, Van Riper, and Ruggles (2019)).

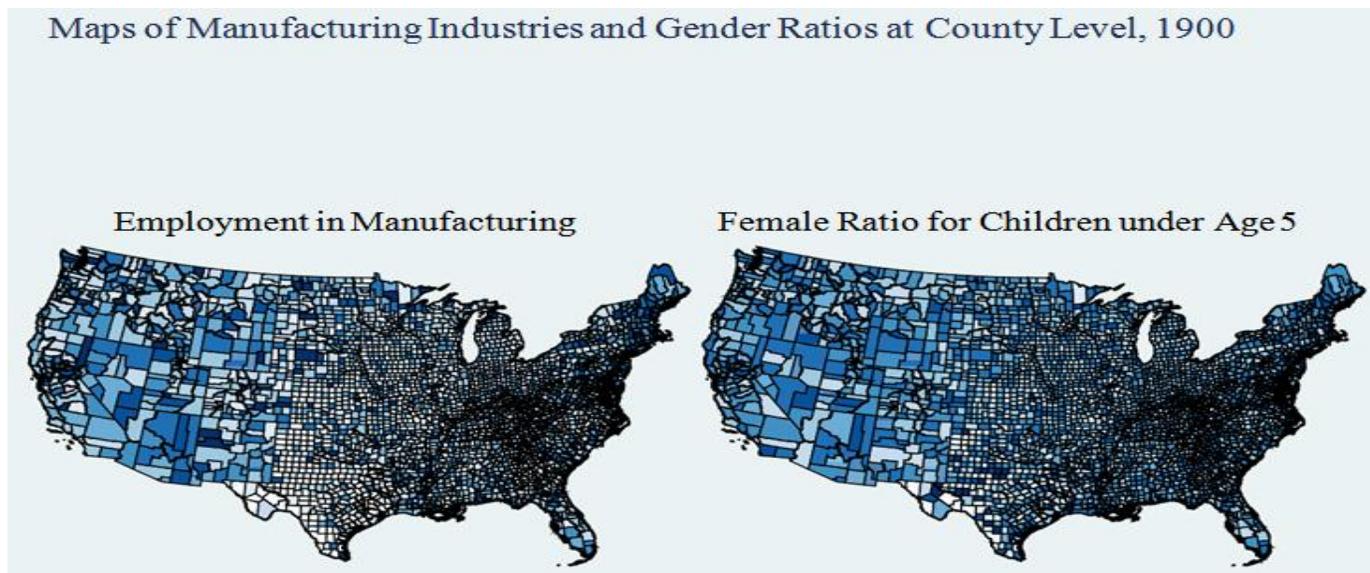

**Figure-1.** Manufacturing growth at the dawn of 20th century.
**Notes:** Full decennial census of 1900 is used. Colors are based on deciles of each variable.

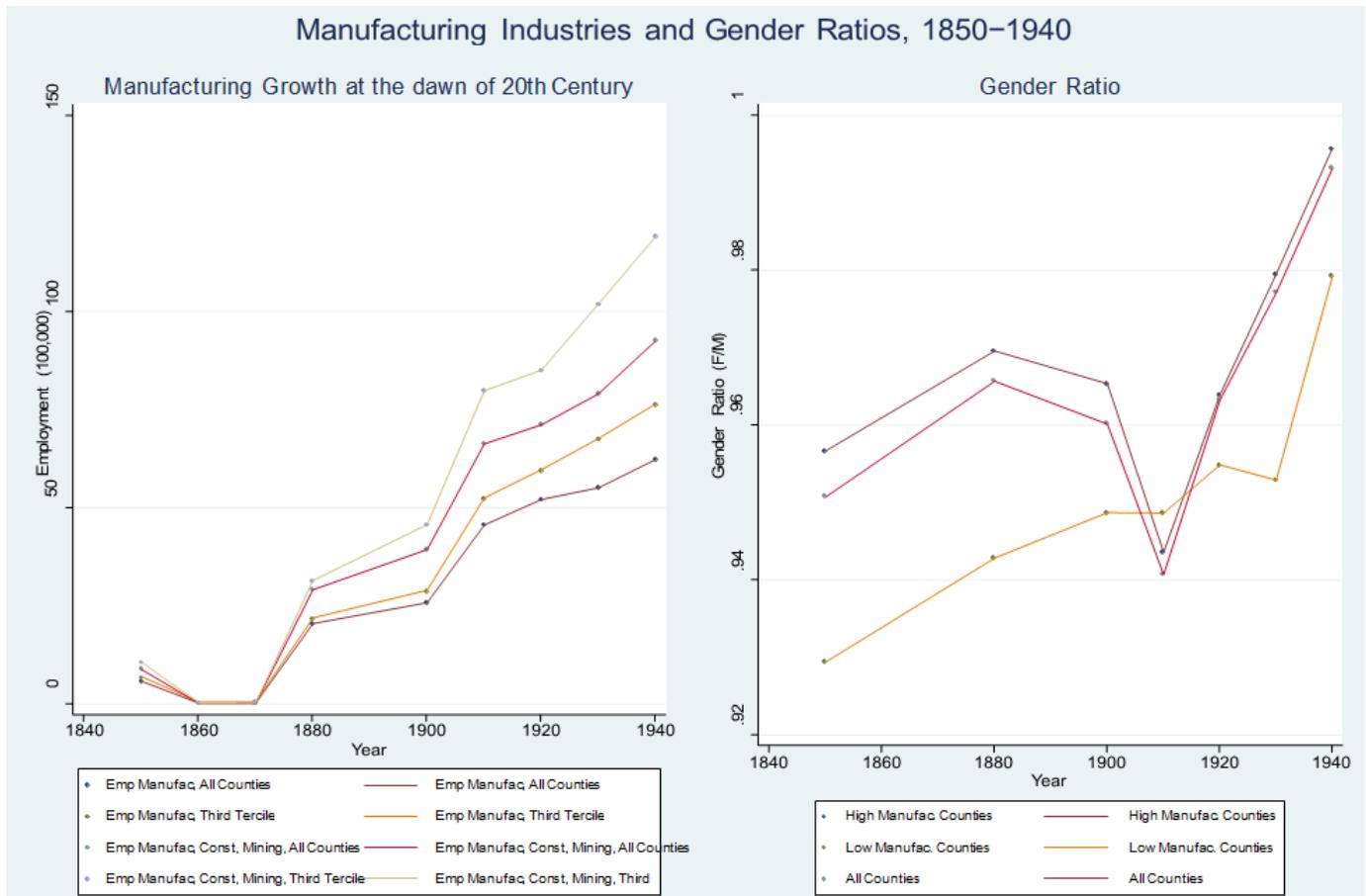

**Figure-2.** Manufacturing Growth at the dawn of 20th Century.
**Notes:** Full Count Decennial Censuses are used. First and third terciles of percentage of manufacturing in all industries at county level is used to define high/low manufacturing intensity.

We merge the employment in manufacturing data with household level data based on county and year. In Table 4 an illustrative summary statistics is presented. To account for the spatial variation in log of employment in manufacturing and female ratio for children under age 5, the differences between the selected variables for counties with high and low employment in manufacturing are reported in the last rows spanning three descriptive years. although in 1880, the female newborns accounted for 49.31% of total children under age 5 in high manufacturing counties and 49.58% in low manufacturing ones (a -0.26% difference) it increased in 1910 to 49.53% and 49.06%, respectively, a difference of 0.47%. Meanwhile the log of employment in manufacturing exhibited a difference of 7.7 and 8.1, respectively.

The use of female ratio in the population as a metric for infant and fetal death rates is validated using some contemporary datasets. Birth Data Files are linked to Mortality Multiple Cause Files for the years between 1968–2002, extracted from Health Statistics (2002a) and aggregated at county level to calculate death rates for infants of different ages. Another source of data to validate this metric, is the Prenatal Mortality Data Files for the years 1996–2002 extracted from Health Statistics (2002b). The latter dataset combines three separate data files: fetal death datasets, linked birth datasets and infant death dataset. It covers reported fetal death up to 20 weeks of gestation, between 20 and 38 weeks of gestation, and finally infant deaths. Each outcome is aggregated into county level and linked to the county birth files based on county, year, and month of occurrence. To transform the coefficients into elasticity, log female birth at each county–year is regressed on log infant death and the results are reported in Table 1 using year fixed effects. The estimates are strongly significant and elasticities are expectedly high. For deaths caused by health issues among infants up to 10 days old their elasticity is 88% with $R^2$ of 80%. Using prenatal mortality data files, this elasticity is about 93% for infant deaths and 95% for fetal death up to 20 weeks of gestation. For all outcomes of prenatal mortality rates the elasticity is 74% and the $R^2$ is 79%. Significant and persistently high elasticities and quite large R–Squared confirm the potential of the gender ratio, as the selected metric, and fetal and infant death rates, in all periods of prenatal development and infancy up to one year old. In Table 2 the same estimates for different subgroups are demonstrated. The elasticities for whites and blacks are very close to each other for infants up to one year old, 94% for whites and 97% for blacks. Similar results are found for infant deaths using prenatal mortality dataset, 74% and 86% for whites and blacks, respectively. Fetal and infant deaths occurred among low educated and high educated mothers show almost the same elasticities, 85% with R squared equals 89%.

The size of the manufacturing industries at county level is the proxy for pollution as the variable of interest. However, the most recent datasets reporting concentrations of pollutants are presented by Environmental Protection Agency (EPA) from 1980 onwards[1]. Although the emission intensities of industrial sectors over time has changed but we might expect that during 20th century, plans and equipment efficiencies to have increased so that any correlation between pollutants and the size of manufacturing offers a downside estimation of its correlation at the beginning of the century.

Two sources of data are used to compute the selected measures of employment at state level: first, Current Population Survey (CPS) between the years 1980–2017. The integrated datasets have been extracted from Flood, King, Ruggles, and Warren (2017). Using person weights transforms the constructed measures of employment into state and national representative statistics. Percentage of manufacturing for the CPS datasets is the percentage of all respondents who are employed at manufacturing at each state–year cell to the total population of the same cell. Second source is Quarterly Census of Employment and Wages (QCEW) extracted from Bureau of Labor Statistics which reports employment at all industries, disaggregated by year–state–sector, for the years 1990–2016. We merge the EPA-extracted pollution data and QCEW-extracted manufacturing data based on year and county.

---

[1] data source: E. P. A. (2017).

**Table-1.** Validating measures gender ratio as a proxy for prenatal mortality and infant mortality for different periods of gestation and different ages of the infants.

| Dataset: | Multiple Cause Deaths (1968-2002) | | | | Prenatal Mortality (1996-2002) | | | |
|---|---|---|---|---|---|---|---|---|
| | All | 0-10 days | 10-30 days | 1-12 months | All | Infant | 0-20 weeks gestation | 20-38 weeks gestation |
| Dep. Var.: | | | | | | | | |
| Log(Female Birth) | b/t | b/t | b/t | b/t | b/t | b/t | b/t | b/t |
| Log(Infant Death) | 0.906*** | 0.880*** | 0.957*** | 0.934*** | 0.739*** | 0.926*** | 0.946*** | 0.236*** |
| | (519.66) | (347.33) | (225.23) | (242.90) | (47.19) | (101.08) | (109.82) | (15.67) |
| Year FE | Yes | Yes | Yes | Yes | Yes | Yes | Yes | Yes |
| $R^2$ | 0.871 | 0.779 | 0.684 | 0.693 | 0.789 | 0.945 | 0.953 | 0.307 |
| Observations | 64902 | 58981 | 42849 | 47780 | 602 | 602 | 602 | 563 |

**Note:** $t$ statistics in parentheses.
\* $p < 0.05$, \*\* $p < 0.01$, \*\*\* $p < 0.001$.

**Table-2.** Validating measures gender ratio as a proxy for infant mortality using county level data four different subgroups.

| Dataset: | Multiple Cause Deaths (1968-2002) | | | Prenatal Mortality (1996-2002) | | | | |
|---|---|---|---|---|---|---|---|---|
| | Whites | Blacks | Other Races | Whites | Blacks | Other Races | Low Educ | High Educ |
| Dep. Var.: | | | | | | | | |
| Log(Female Birth) | b/t | b/t | b/t | b/t | b/t | b/t | b/t | b/t |
| Log(Infant Death) | 0.935*** | 0.986*** | 0.648*** | 0.737*** | 0.861*** | 0.787*** | 0.855*** | 0.846*** |
| | (495.93) | (157.91) | (71.10) | (43.18) | (73.07) | (48.21) | (70.28) | (71.98) |
| Year FE | Yes | Yes | Yes | Yes | Yes | Yes | Yes | Yes |
| $R^2$ | 0.864 | 0.640 | 0.643 | 0.758 | 0.901 | 0.807 | 0.893 | 0.897 |
| Observations | 62342 | 17935 | 13122 | 602 | 595 | 567 | 602 | 602 |

**Note:** $t$ statistics in parentheses.
\* $p < 0.05$, \*\* $p < 0.01$, \*\*\* $p < 0.001$.

**Table-3.** Validating measures of employment in manufacturing as a proxy for pollution using state level data.

| Dataset: | CPS (1980-2017) | | QCEW (1990-2016) | |
|---|---|---|---|---|
| | Log(Emp Manufac) | %Manufac | Log(EmpManufac) | %Manufac |
| Log(CO) | 0.201*** | 0.388*** | 0.041 | 0.266*** |
| Log(Pm10) | 0.244*** | 0.306*** | 0.271*** | 0.271*** |
| Log(Ozon) | 0.264*** | 0.291*** | 0.278*** | 0.291*** |
| Log(SO2) | 0.286*** | 0.422*** | 0.173*** | 0.330*** |
| Log(Lead) | 0.274*** | 0.356*** | 0.245*** | 0.338*** |
| Observations | 1989 | 1989 | 1938 | 1938 |

**Note:** The pollutants include, in order of top to bottom,: Carbon Monoxide, Particulate Matters of the Less than 10m, Ozon, Sulfur Dioxide, lead.
\* $p < 0.05$, \*\* $p < 0.01$, \*\*\* $p < 0.001$.

**Table-4.** Summary statistics for log of manufacturing and number of girls and boys up to age 10 at county level.

| Variables: | (1880) | (1910) | (1930) |
|---|---|---|---|
| High Manufacturing Counties: Ln(Emp Manufact) | 10.00 | 10.85 | 11.15 |
| %Manufact to Population | 4.244 | 5.494 | 5.508 |
| Ln(Male, Under 10) | 11.37 | 11.71 | 11.88 |
| Ln(Female, Under 10) | 11.34 | 11.69 | 11.85 |
| %Female | 49.31 | 49.53 | 49.30 |
| Low Manufacturing Counties: Ln(Emp Manufact) | 2.277 | 2.743 | 3.894 |
| %Manufact to Population | 0.486 | 0.351 | 0.722 |
| Ln(Male, Under 10) | 5.345 | 6.555 | 6.948 |
| Ln(Female, Under 10) | 5.165 | 6.518 | 6.893 |
| %Female | 49.58 | 49.06 | 48.63 |
| Differences (High - Low): | | | |
| Δ Ln(Emp Manufact) | 7.725 | 8.109 | 7.258 |
| Δ %Manufact to Population | 3.758 | 5.143 | 4.786 |
| Δ %Female | -0.263 | 0.471 | 0.663 |
| Number of Counties | 251 | 282 | 282 |
| Mean coefficients | | | |

Table 3 demonstrates the correlations between log of some of selected pollutants and log of employment in manufacturing, and percentage of manufacturing to total population at state level. The percentage of manufacturing (3, fourth column) is the number of employed people in manufacturing divided by the number of people employed in all industries at each state–year cell. The estimated correlations are strongly significant and persistently robust for different pollutants, different measures of the size of the manufacturing industries, and for both datasets. The elasticity of the correlation between the production of sulfur dioxide and employment in manufacturing is roughly 29%, the same correlation with ozone is 27%, lead 28% and carbon monoxide 20%. Using QCEW, one percentage point increase in carbon monoxide is associated with 27% increase in the share of manufacturing to total industries. Such robust associations and statistically significant correlations advocate our use of manufacturing as a proxy for pollution.

## 4. The Empirical Model and Main Results

In order to assess the causal relationship between air pollution induced by industrial activities on fetal death and infant mortality, I apply the following model using full count decennial censuses between 1850 until 1940:

$$y_{it}^c = \beta_0 + \beta_1 Z_{it}^1 + \beta_2 Z_c^2 + \alpha \tilde{X}_{c,t-1} + \mu_t + \phi_c + \nu_{ict} \qquad (1)$$

Since neither infant mortality nor measures of air pollution are available during 19th century and first decades of 20th century, I used two proxy variables for both dependent and independent variables. $y_{it}^c$ is the probability that a mother $i$ residing at county $c$ at decennial year $t$, who has at least one child, and the age of her eldest child is below a threshold (5, 10, or 15 years of age), has more daughters than sons (if any). This variable captures the excess of females to males at household level, my proxy for infant mortality. I include in $Z_{it}^1$ some covariates of mother (quadratic function of age), and socioeconomic class of the household head (Socioeconomic Index), and in $Z_c^2$ a measure of county area. $\tilde{X}_{c,t-1}$ is the variable of interest which has been proxied by lagged values of the size and relative importance of manufacturing industries at county level using different definitions in different scenarios: percentage of people employed in manufacturing to the total population, percentage of people employed in manufacturing, construction, and mining industries to the total population, log of people employed in manufacturing, or manufacturing, construction, and mining industries. $\mu_t$ and $\phi_c$ includes year fixed effects and county fixed effects and $\nu_{ict}$ is a disturbance term. Therefore, $\alpha$ not only captures the effect of an increase in industrial air pollution on fetal death and infant mortality but also could be used to interpret the industrial pollution effect on gender ratio in the population.

To construct the lagged values for the proxy variable of pollution, the following function has been applied:

$$\tilde{X}_{c,t-1} = \alpha(G, C_t)\tilde{X}_{c,t} + (1 - \alpha(G, C_t))X_{c,t-1} \qquad (2)$$

$$\alpha(G, C_t) = \frac{C_t - C_{t-1} - G/2}{C_t - C_{t-1}} \qquad (3)$$

Where $\backslash C_t$ is the census year t ($t \in$ 1850, 1880, 1900, 1910, 1920, 1930, and 1940) and $G$ is the age group of children ($G \in$ 5, 10, and 15\$) and $G/2$ is the midpoint of age group since in short period of times considered in each census year the fertility rates are roughly uniform. The main reason to construct this function is to capture as much closely as possible the features of the environments in which mothers spent their pregnancy rather than just current period that due to large increases in business and industry could bias the estimates. Another reason is the difference among intervals of decennial files available. For example, applying variables of 1850 as a proxy for families at 1880 reveals asymmetric facts compared to the case when we consistently use variables of 1880 as a proxy for 1900. The third reason is the difference among child groups chosen. The mothers who have children of at most 15 years old at 1910 are more probable to have experience their pregnancy close to 1900 compared to mothers of children at most 5 that might have had their conception closer to 1910.

**Table-5.** Family level analysis: Employment of manufacturing and excess female offspring for mothers with children of at most 5 years old.

|  | (1) | (2) | (3) | (4) | (5) | (6) |
|---|---|---|---|---|---|---|
| Log(Emp Manufac.) | 0.004*** (0.000) | 0.004*** (0.000) | 0.003*** (0.000) |  |  |  |
| Log(Emp Manufac., Coal & Mining) |  |  |  | 0.004*** (0.000) | 0.005*** (0.000) | 0.003*** (0.000) |
| SocioEcon Index(1-10) |  | 0.005*** (0.000) | 0.005*** (0.000) |  | 0.005*** (0.000) | 0.005*** (0.000) |
| Log(County Area) |  | -0.005*** (0.001) | -0.004*** (0.001) |  | -0.006*** (0.001) | -0.004*** (0.001) |
| Age Quad. | No | Yes | Yes | No | Yes | Yes |
| Race Dummies | No | Yes | Yes | No | Yes | Yes |
| Year FE | No | No | Yes | No | No | Yes |
| Number of Cases | 17,043,250 | 17,043,250 | 17,043,250 | 17,043,502 | 17,043,502 | 17,043,502 |

**Note:** Standard errors in parentheses are clustered on county.
* $p < .10$, ** $p < .05$, *** $p < .01$.

**Table-6.** Family level analysis: Employment of manufacturing and excess female offspring for mothers with children of at most 10 years old.

|  | (1) | (2) | (3) | (4) | (5) | (6) |
|---|---|---|---|---|---|---|
| Log(Emp Manufac.) | 0.001*** (0.000) | 0.003*** (0.000) | 0.003*** (0.000) |  |  |  |
| Log(Emp Manufac., Coal & Mining) |  |  |  | 0.001*** (0.000) | 0.003*** (0.000) | 0.003*** (0.000) |
| SocioEcon Index(1-10) |  | 0.003*** (0.000) | 0.003*** (0.000) |  | 0.003*** (0.000) | 0.003*** (0.000) |
| Log(County Area) |  | -0.005*** (0.001) | -0.004*** (0.001) |  | -0.005*** (0.001) | -0.005*** (0.001) |
| Age Quad. | No | Yes | Yes | No | Yes | Yes |
| Race Dummies | No | Yes | Yes | No | Yes | Yes |
| Year FE | No | No | Yes | No | No | Yes |
| Number of Cases | 30,950,895 | 30,950,895 | 30,950,895 | 30,951,355 | 30,951,355 | 30,951,355 |

**Note:** Standard errors in parenthese sare clustered on county.
* $p < .10$, ** $p < .05$, *** $p < .01$.

External shocks at placenta during pregnancy will have more effects on male embryos. To capture this latent effect during pregnancy, we divide the sample into three arbitrary subsamples based on the age of the eldest child present at the household residing with the mother. The outcome variable will be computed for mothers, who has children, and for whom the eldest child is present at the household, and his or her age is less than or equal to 5, 10, and 15 years. The results of one for two measures of independent variable (i.e. logarithm of employment in manufacturing, and percentage of employment in manufacturing to the total population at county levels), and each of the three subsamples are shown in Table 5, 6, 7 using log of employment, and Table 8, 9, and 10 using the percentage of manufacturing employment to the total population as the proxy for pollution.

**Table-7.** Family level analysis: Employment of manufacturing and excess female offspring for mothers with children of at most 15 years old.

|  | (1) | (2) | (3) | (4) | (5) | (6) |
|---|---|---|---|---|---|---|
| Log(Emp Manufac.) | 0.001*** | 0.003*** | 0.003*** |  |  |  |
|  | (0.000) | (0.000) | (0.000) |  |  |  |
| Log(Emp Manufac., Coal & Mining) |  |  |  | 0.001*** | 0.003*** | 0.003*** |
|  |  |  |  | (0.000) | (0.000) | (0.000) |
| SocioEcon Index(1-10) |  | 0.003*** | 0.003*** |  | 0.003*** | 0.003*** |
|  |  | (0.000) | (0.000) |  | (0.000) | (0.000) |
| Log(County Area) |  | -0.005*** | -0.005*** |  | -0.005*** | -0.005*** |
|  |  | (0.001) | (0.001) |  | (0.001) | (0.001) |
| Age Quad. | No | Yes | Yes | No | Yes | Yes |
| Race Dummies | No | Yes | Yes | No | Yes | Yes |
| Year FE | No | No | Yes | No | No | Yes |
| Number of Cases | 45,312,147 | 45,312,147 | 45,312,147 | 45,312,756 | 45,312,756 | 45,312,756 |
|  | (1) | (2) | (3) | (4) | (5) | (6) |
| Log(Emp Manufac.) | 0.001*** | 0.003*** | 0.003*** |  |  |  |
|  | (0.000) | (0.000) | (0.000) |  |  |  |
| Log(Emp Manufac., Coal & Mining) |  |  |  | 0.001*** | 0.003*** | 0.003*** |
|  |  |  |  | (0.000) | (0.000) | (0.000) |
| SocioEcon Index(1-10) |  | 0.003*** | 0.003*** |  | 0.003*** | 0.003*** |
|  |  | (0.000) | (0.000) |  | (0.000) | (0.000) |
| Log(County Area) |  | -0.005*** | -0.005*** |  | -0.005*** | -0.005*** |
|  |  | (0.001) | (0.001) |  | (0.001) | (0.001) |
| Age Quad. | No | Yes | Yes | No | Yes | Yes |
| Race Dummies | No | Yes | Yes | No | Yes | Yes |
| Year FE | No | No | Yes | No | No | Yes |
| Number of Cases | 45,312,147 | 45,312,147 | 45,312,147 | 45,312,756 | 45,312,756 | 45,312,756 |

**Note:** Standard errors in parentheses are clustered on county.
* $p < .10$, ** $p < .05$, *** $p < .01$.

Among mothers with children up to age 5, a 1% increase in the number of people employed in manufacturing at the county of residence is associated with an increase in the probability that the mother has more daughters than sons by 0.3%. This probability is quite robust in other specifications and the same using log of employment in manufacturing, construction and mining as the proxy variable, and among mothers with children up to age 10 and 15 as well. The vast county area can act as a poppet valve to reduce the pressure caused by manufacturing plants and disperse the pollutants in a wider horizon. Thus, it is expected that the degree of expansiveness of a county lowers the negative health effects of the pollution including fetal deaths. The inclusion of the above county area and negative statistically significant coefficients confirm this assumption. While log of employment in manufacturing industries is a measure of total pollution at county level, the percentage of manufacturing employment to the total population is a measure of per person pollution in each county. The probability that the number of daughters exceed the number of sons in a family whose mother's eldest child is at most five will increase by 3.7% if the share of people employed in manufacturing industries to the total population of the county increase by one percentage point. If people employed in construction and mining be added to this ratio, the probability will be 2.9%. These probabilities will increase for the subsamples of mothers with children of at most 10 and 15. 1% increase in the share of manufacturing employment will increase the probability that a mother has more daughters than sons by roughly 4.7 and 5.7 percentage points, respectively. The coefficients of interest in all specifications are statistically and economically significant at 1% level. The sex ratio of most species tends to be 1:1 as stated by well-known fisher's principal (Fisher, 1999).[2] However, some skewness is probable in this principle as a response to some environmental factors in order to maximize the chances of survival and reproduction of the species. Male embryos are generally weaker and are more probable to die whether in prenatal development or as an infant. Moreover, a male member must be strong and healthy enough to support a family and attract a mate while a typical female, regardless of its health status, has higher chances to find a male for a mate. Negative environmental shocks are being interpreted by reproductive system of the mother that male embryos are more in danger of being vanished without an offspring. Due to higher chances of survival and reproduction of females, during such negative shocks the sex ratio will be skewed towards more females than males among newborns. Therefore, a 3% increase in the female of the newborns is a signal for not only fetal and infant deaths, but also a signal of infants that could have been died or at least were more likely endangered had an XY–chromosomes matched at the time of conception rather than an XX. As a reminder, labile sex determination is being favored among humans by evolution since environmental factors had not had any significant effect on their survival and that is why an environmental–driven sex determination among humans is not something expected to be observed, a rare phenomenon occurred at special circumstances.

---

[2] Although with some selective deviation (as in peafowl birds (Pike & Petrie, 2005) or sex transforms in dichogamies (Warner, 1975) or some temperature dependent sex determination (female and males of American alligator hatch their egg at different temperatures).

**Table-8.** Family level analysis: Percentage of manufacturing and excess female offspring for mothers with children of at most 5 years old.

|  | (1) | (2) | (3) | (4) | (5) | (6) |
|---|---|---|---|---|---|---|
| %Manufacturing | 0.043*** (0.005) | 0.038*** (0.012) | 0.037*** (0.010) |  |  |  |
| %Manufac., Coal & Mining |  |  |  | 0.044*** (0.004) | 0.041*** (0.010) | 0.029*** (0.009) |
| SocioEcon Index(1-10) |  | 0.005*** (0.000) | 0.005*** (0.000) |  | 0.005*** (0.000) | 0.005*** (0.000) |
| Log(County Area) |  | -0.002 (0.001) | -0.002* (0.001) |  | -0.002* (0.001) | -0.002 (0.001) |
| Age Quad. | No | Yes | Yes | No | Yes | Yes |
| Race Dummies | No | Yes | Yes | No | Yes | Yes |
| Year FE | No | No | Yes | No | No | Yes |
| Number of Cases | 17,043,558 | 17,043,558 | 17,043,558 | 17,043,558 | 17,043,558 | 17,043,558 |

**Note:** Standard errors in parentheses are clustered on county.
\* $p < .10$, \*\* $p < .05$, \*\*\* $p < .01$.

**Table-9.** Family level analysis: Percentage of manufacturing and excess female offspring for mothers with children of at most 10 years old.

|  | (1) | (2) | (3) | (4) | (5) | (6) |
|---|---|---|---|---|---|---|
| %Manufacturing | 0.018*** (0.004) | 0.036*** (0.007) | 0.047*** (0.009) |  |  |  |
| %Manufac., Coal & Mining |  |  |  | 0.016*** (0.003) | 0.032*** (0.006) | 0.044*** (0.007) |
| SocioEcon Index(1-10) |  | 0.004*** (0.000) | 0.004*** (0.000) |  | 0.004*** (0.000) | 0.004*** (0.000) |
| Log(County Area) |  | -0.002** (0.001) | -0.003** (0.001) |  | -0.003** (0.001) | -0.003** (0.001) |
| Age Quad. | No | Yes | Yes | No | Yes | Yes |
| Race Dummies | No | Yes | Yes | No | Yes | Yes |
| Year FE | No | No | Yes | No | No | Yes |
| Number of Cases | 30,951,459 | 30,951,459 | 30,951,459 | 30,951,459 | 30,951,459 | 30,951,459 |

**Note:** Standard errors in parentheses are clustered on county.
\* $p < .10$, \*\* $p < .05$, \*\*\* $p < .01$.

**Table-10.** Family level analysis: Percentage of manufacturing and excess female offspring for mothers with children of at most 15 years old.

|  | (1) | (2) | (3) | (4) | (5) | (6) |
|---|---|---|---|---|---|---|
| %Manufacturing | 0.008** (0.003) | 0.027*** (0.008) | 0.057*** (0.010) |  |  |  |
| %Manufac., Coal & Mining |  |  |  | 0.003 (0.003) | 0.022*** (0.007) | 0.055*** (0.008) |
| SocioEcon Index(1-10) |  | 0.003*** (0.000) | 0.003*** (0.000) |  | 0.003*** (0.000) | 0.003*** (0.000) |
| Log(County Area) |  | -0.003* (0.002) | -0.003** (0.001) |  | -0.003 (0.002) | -0.003** (0.001) |
| Age Quad. | No | Yes | Yes | No | Yes | Yes |
| Race Dummies | No | Yes | Yes | No | Yes | Yes |
| Year FE | No | No | Yes | No | No | Yes |
| Number of Cases | 45,312,890 | 45,312,890 | 45,312,890 | 45,312,890 | 45,312,890 | 45,312,890 |

**Note:** Standard errors in parentheses are clustered on county.
\* $p < .10$, \*\* $p < .05$, \*\*\* $p < .01$.

## 5. Robustness Checks

### 5.1. Gender Preference among Parents

In their paper, Dahl and Moretti (2008) show that US parents favor sons over daughters. They use 1960–2000 census data and show that the number of children is significantly higher for the families whose firstborn child is a girl. Moreover, a divorce is the more likely occurrence among families with one daughter and after divorce, fathers are more likely to take custody of sons rather than daughters. Furthermore, child gender in utero, revealed by ultrasound tests, has significant effect on shotgun marriages. Although the range of time is quiet contemporary but the cultural factor has high inertia and is passed from one generation to another easily. Therefore, I expect the same gender preference among families even a couple of decades earlier, the time span of this research. Based on these lines of reasoning, families with only one child are more likely to be families with a boy and families are more likely to move to the second and higher children in case they have a girl as their first child. If one–child families is associated with one–boy families, then in the subsample of mothers with only one child, the coefficients of interest must suffer a downward bias to zero, implying that a more reliable subsample is families with at least two children. To address this potential bias, I run the Equation 1 on subsamples of mothers with only one child, only two children, and three or more children for different age groups. As shown in Table 11, for children of up to age 5, the coefficient increases by 0.1% when we move from the subsample of families with only two children to the families with three or more children while the coefficient is insignificant and negligible for the subsample of families with only one child. The magnitude of this difference is larger for children up to age 10 and 15. While for the families with only two children whose eldest child is 10 and 15, the coefficient on log of employment in manufacturing is 0.1% and 0.2% accordingly. These coefficients increase to 0.4% and 0.5% respectively, for families with three and more children.

Although it's not obvious that families with more than one child had a girl for the first child but the association between larger families and higher probabilities of having more girls than boys confirms my concern about the downward bias due to gender preferences among parents as a driver force for fertility decisions. Using log of employment in manufacturing, construction and mining as a proxy variable, presented in Table 12 uncover the same information. The effects of an increase in log of employment in manufacturing, construction and mining will increase the probability that a mother has more daughters than sons by higher magnitudes for mothers with more children. Table 13 and Table 14 present the results for other proxy variables. A 1% increase in share of manufacturing to total population among children up to age 10 will increase the probability of more females been born to a mother by 1.2% for mothers who have only one child while for mothers with only two children, and three or more children the coefficients raises to 2.9% and 9.1% respectively.

### *5.2. Aggregations*

In the next step, I move on to county levels and aggregate household data into county level data for all variables. I then run the regression of the share of females among children on employment in manufacturing controlling for percentage of white residents in the county, percentage of blacks, log of county area, average female labor force participation of the county, and average socioeconomic index of county residents. The results for different age groups are reported in Table 15. A 1% increase in employment of manufacturing is correlated with a 0.1% increase in the share of female in the population of children in different age groups regardless of using any mentioned controls or not. Persistently significant results in Table 16 using percentage of manufacturing as a proxy variable support the previous results at the aggregate level as well. For children of at most five years old, 1% increase in the ratio of manufacturing to total population is associated with an increase of 2.4% female share in the population.

### *5.3. Female Labor Force Participation*

Fogli and Veldkamp (2011) show that the growth of maternal labor force participation is being influenced by the information a woman gathers from the nearby employed mothers. Therefore, the growth of female labor force participation at counties with less female participated in the last period grows slowly compared to counties where more women has participated earlier. Based on this fact, I use percentage of females employed at manufacturing industries at county level as a proxy for the probability of exposure to pollution before pregnancy. If in a certain county more women are willing to participate in the labor force, and I observe their labor force participation at the current period when they already had given birth to their children, I would expect more mothers in this county had been exposed to pollution due to their previous labor force participation. The main problem is that the share of female employment in manufacturing might have some interaction with total size of manufacturing industry in a specific county. If higher size of manufacturing industries encourages more participation of women in the labor force then I would expect an upward bias in the coefficients while if lower size of manufacturing in one county was due to, for example, worse geographical location of the county and so more women have to participate for subsistence of the family, then the share of woman in the manufacturing, which its size is low in this case, will rise and tend to bias estimates towards zero. Therefore, I avoid to interpret the coefficients of this section into causal relations. However, the consistently significant coefficients as presented in Table 17 even by adding county level controls and year fixed effects confirm the positive association between percentage of female employed in manufacturing and ratio of female among children. A 1% increase in the share of female employed in manufacturing industries to the total population is statistically associated with 11.4%, 15.4%, and 18.6% raise in female ratio of children up to age 5, 10, and 15, respectively.

## 6. Discussions

The aim of this paper was to investigate the effects of the sharp rise in industrial pollution on prenatal and infant mortality rates in the late 19th century and the first decades of the 20th century. To overcome the challenge of the scarcity of data in the selected time period, industrial pollution was proxied by some measures of the size and relative importance of manufacturing industries at county level. We first established an empirical context for the appropriateness of our proxies using contemporaneous datasets. Specifically, we showed that employment in manufacturing is indeed associated with higher pollution. Moreover, we demonstrated the fact that sex ratio at birth is a good measure for infant and fetal death. Then, we applied the concentration in manufacturing industries as the proxy for pollution and the sex ratio at birth as the proxy for infant mortality. Using a panel data fixed effect model, we found that the up-rise in pollution during the late nineteenth century and the beginning of the twentieth century was associated with higher rates of infant mortality and fetal death. we found that one thousandth of logarithm of employment in manufacturing is associated with 1–3% increase in the probability of a live birth being female.

The reproductive system of the mother responds to external stressors and assigns different probabilities to male and female in order to maximize the likelihood of survival and reproduction. Since female adults, regardless of their health status, are more likely to mate compared to male adults and moreover, male embryos are more susceptible to lower health quality of mother and environmental shocks to placenta, the assigned probabilities at the time of conception will deviate to the benefit of females. This fact is the intuition for the choice of gender ratio for fetal and infant deaths.

**Table-11.** Examining the effect of gender preferences at family level analysis: Excess female offspring for mothers and employment of manufacturing.

|  | Only 1 Child | | | Only 2 Child | | | More than 3 Children | | |
|---|---|---|---|---|---|---|---|---|---|
|  | Up to Age 5 | Up to Age 10 | Up to Age 15 | Up to Age 5 | Up to Age 10 | Up to Age 15 | Up to Age 5 | Up to Age 10 | Up to Age 15 |
| Log(Emp Manufac.) | 0.000 | 0.001** | 0.002*** | 0.001*** | 0.001*** | 0.002*** | 0.002*** | 0.004*** | 0.005*** |
|  | (0.000) | (0.000) | (0.000) | (0.000) | (0.000) | (0.000) | (0.001) | (0.000) | (0.000) |
| SocioEcon Index(1-10) | -0.000 | 0.001*** | 0.002*** | -0.000 | 0.000 | 0.001*** | 0.002*** | 0.005*** | 0.006*** |
|  | (0.000) | (0.000) | (0.000) | (0.000) | (0.000) | (0.000) | (0.000) | (0.000) | (0.000) |
| Log(County Area) | -0.001 | -0.002*** | -0.003*** | -0.003*** | -0.002*** | -0.003*** | -0.005*** | -0.004*** | -0.003*** |
|  | (0.001) | (0.001) | (0.001) | (0.001) | (0.001) | (0.001) | (0.001) | (0.001) | (0.001) |
| Age Quad. | Yes | Yes | Yes | Yes | Yes | Yes | Yes | Yes | Yes |
| Race Dummies | Yes | Yes | Yes | Yes | Yes | Yes | Yes | Yes | Yes |
| Year FE | Yes | Yes | Yes | Yes | Yes | Yes | Yes | Yes | Yes |
| Number of Cases | 10,098,188 | 13,565,384 | 16,741,763 | 5,142,617 | 9,211,929 | 12,492,967 | 1,802,445 | 8,173,582 | 16,077,417 |

**Note:** Standard errors in parentheses are clustered on county.

$^* p < .10, ^{**} p < .05, ^{***} p < .01.$

**Table-12.** Examining the effect of gender preferences at family level analysis: Excess female offspring for mothers and employment of manufacturing, construction, and mining.

|  | Only 1 Child | | | Only 2 Child | | | More than 3 Children | | |
|---|---|---|---|---|---|---|---|---|---|
|  | Up to Age 5 | Up to Age 10 | Up to Age 15 | Up to Age 5 | Up to Age 10 | Up to Age 15 | Up to Age 5 | Up to Age 10 | Up to Age 15 |
| Log(Emp Manufac., Coal & Mining) | 0.000 | 0.001** | 0.002*** | 0.001*** | 0.001*** | 0.002*** | 0.002*** | 0.005*** | 0.005*** |
|  | (0.000) | (0.000) | (0.000) | (0.001) | (0.000) | (0.000) | (0.001) | (0.000) | (0.000) |
| SocioEcon Index(1-10) | -0.000 | 0.001*** | 0.002*** | -0.000 | 0.000 | 0.001*** | 0.002*** | 0.005*** | 0.006*** |
|  | (0.000) | (0.000) | (0.000) | (0.000) | (0.000) | (0.000) | (0.000) | (0.000) | (0.000) |
| Log(County Area) | -0.001 | -0.002*** | -0.003*** | -0.003*** | -0.002*** | -0.003*** | -0.005*** | -0.004*** | -0.004*** |
|  | (0.001) | (0.001) | (0.001) | (0.001) | (0.001) | (0.001) | (0.001) | (0.001) | (0.001) |
| Age Quad. | Yes | Yes | Yes | Yes | Yes | Yes | Yes | Yes | Yes |
| Race Dummies | Yes | Yes | Yes | Yes | Yes | Yes | Yes | Yes | Yes |
| Year FE | Yes | Yes | Yes | Yes | Yes | Yes | Yes | Yes | Yes |
| Number of Cases | 10,098,328 | 13,565,544 | 16,741,940 | 5,142,701 | 9,212,063 | 12,493,122 | 1,802,473 | 8,173,748 | 16,077,694 |

**Note:** Standard errors in parentheses are clustered on county.

$^* p < .10, ^{**} p < .05, ^{***} p < .01.$

**Table-13.** Examining the effect of gender preferences at family level analysis: Excess female offspring for mothers and percentage of manufacturing.

|  | Only 1 Child | | | Only 2 Child | | | More than 3 Children | | |
| --- | --- | --- | --- | --- | --- | --- | --- | --- | --- |
|  | Up to Age 5 | Up to Age 10 | Up to Age 15 | Up to Age 5 | Up to Age 10 | Up to Age 15 | Up to Age 5 | Up to Age 10 | Up to Age 15 |
| %Manufacturing | 0.006 | 0.012* | 0.031*** | 0.025** | 0.029*** | 0.036*** | 0.038** | 0.091*** | 0.097*** |
|  | (0.007) | (0.007) | (0.008) | (0.012) | (0.011) | (0.011) | (0.017) | (0.011) | (0.011) |
| SocioEcon Index(1-10) | -0.000 | 0.001*** | 0.002*** | -0.000 | 0.000 | 0.001*** | 0.002*** | 0.005*** | 0.006*** |
|  | (0.000) | (0.000) | (0.000) | (0.000) | (0.000) | (0.000) | (0.000) | (0.000) | (0.000) |
| Log(County Area) | -0.000 | -0.002** | -0.002*** | -0.002* | -0.001 | -0.002** | -0.003** | -0.001 | -0.001 |
|  | (0.001) | (0.001) | (0.001) | (0.001) | (0.001) | (0.001) | (0.001) | (0.001) | (0.002) |
| Age Quad. | Yes | Yes | Yes | Yes | Yes | Yes | Yes | Yes | Yes |
| Race Dummies | Yes | Yes | Yes | Yes | Yes | Yes | Yes | Yes | Yes |
| Year FE | Yes | Yes | Yes | Yes | Yes | Yes | Yes | Yes | Yes |
| Number of Cases | 10,098,360 | 13,565,586 | 16,741,989 | 5,142,717 | 9,212,092 | 12,493,155 | 1,802,48 | 8,173,781 | 16,077,746 |

**Note:** Standard errors in parentheses are clustered on county.
\* $p < .10$, \*\* $p < .05$, \*\*\* $p < .01$.

**Table-14.** Examining the effect of gender preferences at family level analysis: Excess female offspring for mothers and percentage of manufacturing, construction, and mining.

|  | Only 1 Child | | | Only 2 Child | | | More than 3 Children | | |
| --- | --- | --- | --- | --- | --- | --- | --- | --- | --- |
|  | Up to Age 5 | Up to Age 10 | Up to Age 15 | Up to Age 5 | Up to Age 10 | Up to Age 15 | Up to Age 5 | Up to Age 10 | Up to Age 15 |
| %Manufac., Coal & Mining | 0.006 | 0.014** | 0.034*** | 0.018* | 0.025** | 0.032*** | 0.040*** | 0.084*** | 0.089*** |
|  | (0.007) | (0.007) | (0.007) | (0.011) | (0.010) | (0.009) | (0.015) | (0.008) | (0.009) |
| SocioEcon Index(1-10) | -0.000 | 0.001*** | 0.002*** | -0.000 | 0.000 | 0.001*** | 0.002*** | 0.005*** | 0.006*** |
|  | (0.000) | (0.000) | (0.000) | (0.000) | (0.000) | (0.000) | (0.000) | (0.000) | (0.000) |
| Log(County Area) | -0.000 | -0.002** | -0.002*** | -0.002* | -0.001 | -0.002** | -0.003** | -0.001 | -0.001 |
|  | (0.001) | (0.001) | (0.001) | (0.001) | (0.001) | (0.001) | (0.001) | (0.001) | (0.002) |
| Age Quad. | Yes | Yes | Yes | Yes | Yes | Yes | Yes | Yes | Yes |
| Race Dummies | Yes | Yes | Yes | Yes | Yes | Yes | Yes | Yes | Yes |
| Year FE | Yes | Yes | Yes | Yes | Yes | Yes | Yes | Yes | Yes |
| Number of Cases | 10,098,360 | 13,565,586 | 16,741,989 | 5,142,717 | 9,212,092 | 12,493,155 | 1,802,481 | 8,173,781 | 16,077,746 |

**Note:** Standard errors in parentheses are clustered on county.
\* $p < .10$, \*\* $p < .05$, \*\*\* $p < .01$.

**Table-15.** County level analysis: Female ratios among children and manufacturing em- ployment.

|  | Up to Age 5 | | Up to Age 10 | | Up to Age 15 | |
|---|---|---|---|---|---|---|
|  | (1) | (2) | (3) | (4) | (5) | (6) |
| Ln(Emp Manufact) | 0.001*** | 0.001*** | 0.001*** | 0.001*** | 0.001*** | 0.001*** |
|  | (0.000) | (0.000) | (0.000) | (0.000) | (0.000) | (0.000) |
| Ln(Area) |  | −0.001** |  | −0.003*** |  | −0.003*** |
|  |  | (0.001) |  | (0.001) |  | (0.000) |
| Controls | No | Yes | No | Yes | No | Yes |
| Number of Cases | 1,482 | 1,482 | 1,482 | 1,482 | 1,482 | 1,482 |

**Note:** Standard errors in parentheses are clustered on county.

\* $p < .10$, \*\* $p < .05$, \*\*\* $p < .01$.

**Table-16.** County level analysis: Female ratios among children and percentage of manu- facturing employment in total county population.

|  | Up to Age 5 | | Up to Age 10 | | Up to Age 15 | |
|---|---|---|---|---|---|---|
|  | (1) | (2) | (3) | (4) | (5) | (6) |
| %Manufact | 0.028*** | 0.024** | 0.041*** | 0.028*** | 0.056*** | 0.035*** |
|  | (0.010) | (0.012) | (0.008) | (0.010) | (0.007) | (0.008) |
| Ln(Area) |  | −0.001 |  | −0.002*** |  | −0.002*** |
|  |  | (0.001) |  | (0.001) |  | (0.000) |
| Controls | No | Yes | No | Yes | No | Yes |
| Number of Cases | 1,487 | 1,487 | 1,487 | 1,487 | 1,487 | 1,487 |

**Note:** Standard errors in parentheses are clustered on county.

\* $p < .10$, \*\* $p < .05$, \*\*\* $p < .01$.

**Table-17.** County level analysis: Female ratios among children and average female labor force participation in manufacturing sectors.

|  | Up to Age 5 | | Up to Age 10 | | Up to Age 15 | |
|---|---|---|---|---|---|---|
|  | (1) | (2) | (3) | (4) | (5) | (6) |
| %Emp Manufact, Females | 0.130** | 0.114* | 0.205*** | 0.154*** | 0.264*** | 0.186*** |
|  | (0.053) | (0.059) | (0.046) | (0.051) | (0.037) | (0.041) |
| Ln(Area) |  | −0.001* |  | −0.002*** |  | −0.003*** |
|  |  | (0.001) |  | (0.001) |  | (0.000) |
| Controls | No | Yes | No | Yes | No | Yes |
| Number of Cases | 1,485 | 1,485 | 1,485 | 1,485 | 1,485 | 1,485 |

**Note:** Standard errors in parentheses are clustered on county.

\* $p < .10$, \*\* $p < .05$, \*\*\* $p < .01$.

However, it is complicated to quantitatively interpret the results from effects of pollution onto the infant and fetal mortalities for some reasons. First, plants and factories went under unprecedented technological change and this led to different emission intensities, production efficiencies, or even change in raw materials. Therefore, making a bridge between contemporary data of manufacturing and those of late 19$^{th}$ century is likely to produce misleading results. Furthermore, as shown in Table 1 and more noticeably Table 2, although the elasticities between female ratios and prenatal death (and infant mortality) rates are quite high and statistically significant but it varies non–negligibly for different subgroups. The raise of new medicines in the market, constantly changing public environment, other medical fac- tors, and unknown pregnancy makes it even harder to quantitatively report the results as the magnitude of the effect of pollution on fetal deaths. All in all, this paper has shed light to some costs of industrial upswing in US which had been conveyed to embryos and infants and more specifically male infants.